\documentstyle[epsf]{elsart}
\newcommand{\size}{6.5truecm}
\newcommand{\sizetwo}{6.5truecm}
\begin{document}
\begin{frontmatter}
\title{\hspace*{\fill}{\normalsize JYFL 98--11;\hspace*{1em} 
                                   TPR--98--20} \\[1.5ex]
       Sensitivity of electromagnetic spectra to equation
       of state and initial energy density in the Pb+Pb collisions at
       SPS}

\author[Jkl]{Pasi Huovinen},
\author[Jkl]{P.V.~Ruuskanen} and
\author[Regensb]{Josef Sollfrank}
\address[Jkl]{Department of Physics,
              University of Jyv\"askyl\"a, Finland}
\address[Regensb]{Institut f\"ur theoretische Physik, Universit\"at
                  Regensburg, Germany}

\begin{abstract}
We study Pb+Pb collisions at 158 $A$ GeV/$c$ using a hydrodynamical
approach. We test different equations of state (EoSs) and different
initial conditions and show that there are more than one initial state
for each EoS which reproduce the observed hadronic spectra. We also
find that different equations of state favour different freeze-out
temperature. Simultaneously we calculate the thermal dilepton and
photon spectra for each EoS and initial state. We compare the dilepton
mass spectrum to data measured by the CERES collaboration and find
that the differences in spectra obtained using different EoSs and
initial states are not resolvable within the current experimental
resolution. However, at invariant masses over 2 GeV the difference
in the yield due to various initial states is close to an order of
magnitude. We also study the rapidity distribution of lepton pairs
and find that for masses around 800 MeV the shape of the distribution
depends strongly on the EoS.\\
\vspace{5mm}
\noindent PACS number(s): 25.75.-q, 12.38.Mh, 12.40.Ee, 47.75.+f
\end{abstract}
\end{frontmatter}

\section{Introduction}

One aim in the field of relativistic heavy ion collisions is to study the
dynamics of the collisions and try to give reliable estimates about
the achieved energy density with the intention to relate this density
to the phase transition density of the quark-gluon plasma. The
dynamics of these collisions can only be studied by models which are
tested by comparing to the various experimental data. There is a big
effort from the experimental groups to measure as many observables as
possible, like single hadron spectra, correlation functions, photon
spectra and dilepton spectra.

The subset of hadronic spectra gives a reliable estimate only
of the final state at decoupling. Concluding the final state 
from the spectra of one particle species only is not
possible because there is an ambiguity between the freeze-out
temperature and the collective flow velocity \cite{Schnedermann93}.
There are methods to overcome this ambiguity by analyzing
transverse momentum spectra of particles with very different masses
\cite{Kampfer96}, study the coalescence of various small nuclei
and anti-nuclei \cite{Scheibl98} or do an HBT analysis 
\cite{Wiedemann98,Schlei97}.
All these methods help to determine more precisely the final
state of a heavy ion collision but they do not tell anything about the
dynamics producing the final state.

Dynamical models like hydrodynamics or event generators are first of
all tested against the final state. However, reproducing the final
state does not mean that the dynamics before the final state is
reasonable. Electromagnetic probes are emitted during the whole hot
and dense stage of a heavy ion collision. Therefore they are
observables which may help to distinguish between different evolution
scenarios which reproduce the final state. The main goal of our study
is to use electromagnetic spectra to test different hydrodynamical
scenarios which reproduce the final hadronic state in a physically
motivated but still approximative way. 
Similar studies 
\cite{Dumitru93,Srivastava94,Shuryak94,Arbex95,Neumann95,Tarasov96,Srivastava96,Steele96,Chaudhuri97,Hirano97,Cleymans97,Hung97,Sollfrank97a,Sarkar98,Huovinen98}
have already been performed, but mostly for
S+Au collisions. Here we concentrate on Pb+Pb(Au) collisions
and study the effects of the uncertainty in the initial
energy density.

In hydrodynamical models the differences in equation of state (EoS)
can be compensated by the choice of the initial conditions. There are
also several different initial conditions for each EoS which lead to
comparable results concerning single hadron spectra. Here we first
construct two such initial conditions for three different EoSs
and then investigate the power of single photon spectra, low
mass dielectrons, and intermediate mass dimuons to differentiate
between these different cases.

\section{Equation of state}

A necessary ingredient for the hydrodynamic calculation is the equation
of state. The sensitivity of electromagnetic emission to the EoS arises
in two different ways. First, the properties of the quanta of the
matter affect the rate at given  temperature and second, the expansion
timescale and temperature profile depend on the EoS. Typically an EoS
with less degrees of freedom leads to higher temperatures but also to
faster cooling than in the case of large number of degrees of freedom.
To see these differences we investigate three EoSs with different phase
transition temperatures. In the case of EoS~H we assume the matter to
remain in the form of hadrons without plasma formation even at highest
densities. In the case of EoS~D a first order phase transition occurs at
$T_c=200$ MeV at zero net baryon number density and in case A the
phase transition temperature is $T_c=165$ MeV.

In \cite{Sollfrank97a} we have given a detailed description of how
a bag model type equation of state is calculated from given input
parameters. In our case these input parameters are the hadron and
parton degrees of freedom, the bag constant $B$, and a mean field
repulsion characterized by the coupling $K$ of the baryons to the net
baryon number density $\rho_B$. Unlike in our previous 
papers~\cite{Sollfrank97a,Sollfrank97b,Sollfrank97c} we now include
all the hadrons up to 2 GeV mass listed in Particle Data 
Book~\cite{PDB} for constructing the hadronic part of the EoS while
the QGP part is unchanged. We then obtain the values $T_c=165$ (EoS~A)
and 200 MeV (EoS~D) with the mean field parameter $K=450$ MeV and 
$B^{1/4} = 235$ and 264 MeV, respectively. In case of EoS~H we have
$K=450$ MeV.

\section{Initial conditions}

\begin{figure}
\begin{center}
  \begin{minipage}[t]{\size}
        \epsfxsize \size \epsfbox{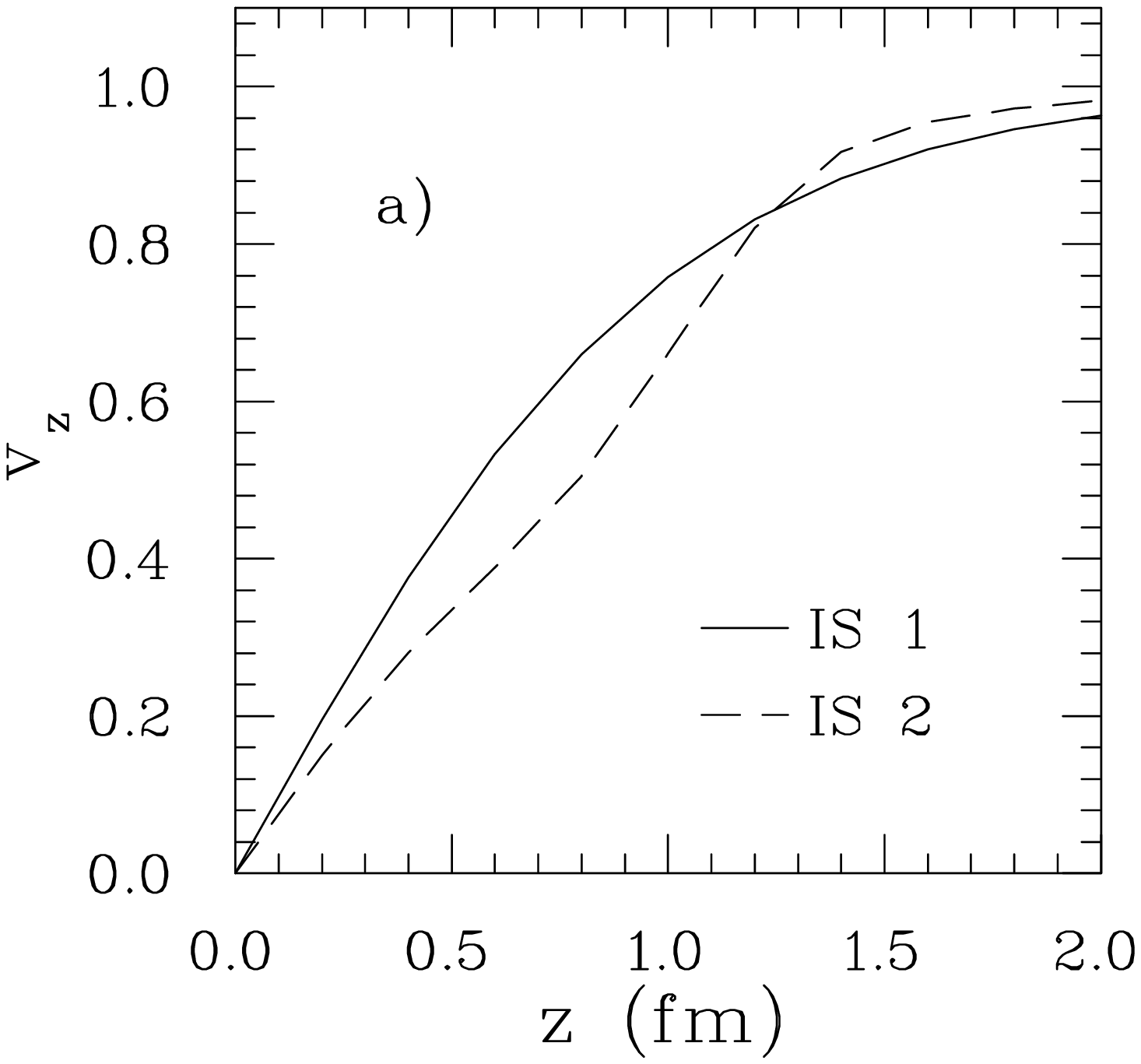}
        \hfill
  \end{minipage}
   \hfill
  \begin{minipage}[t]{\size}
        \epsfxsize \size \epsfbox{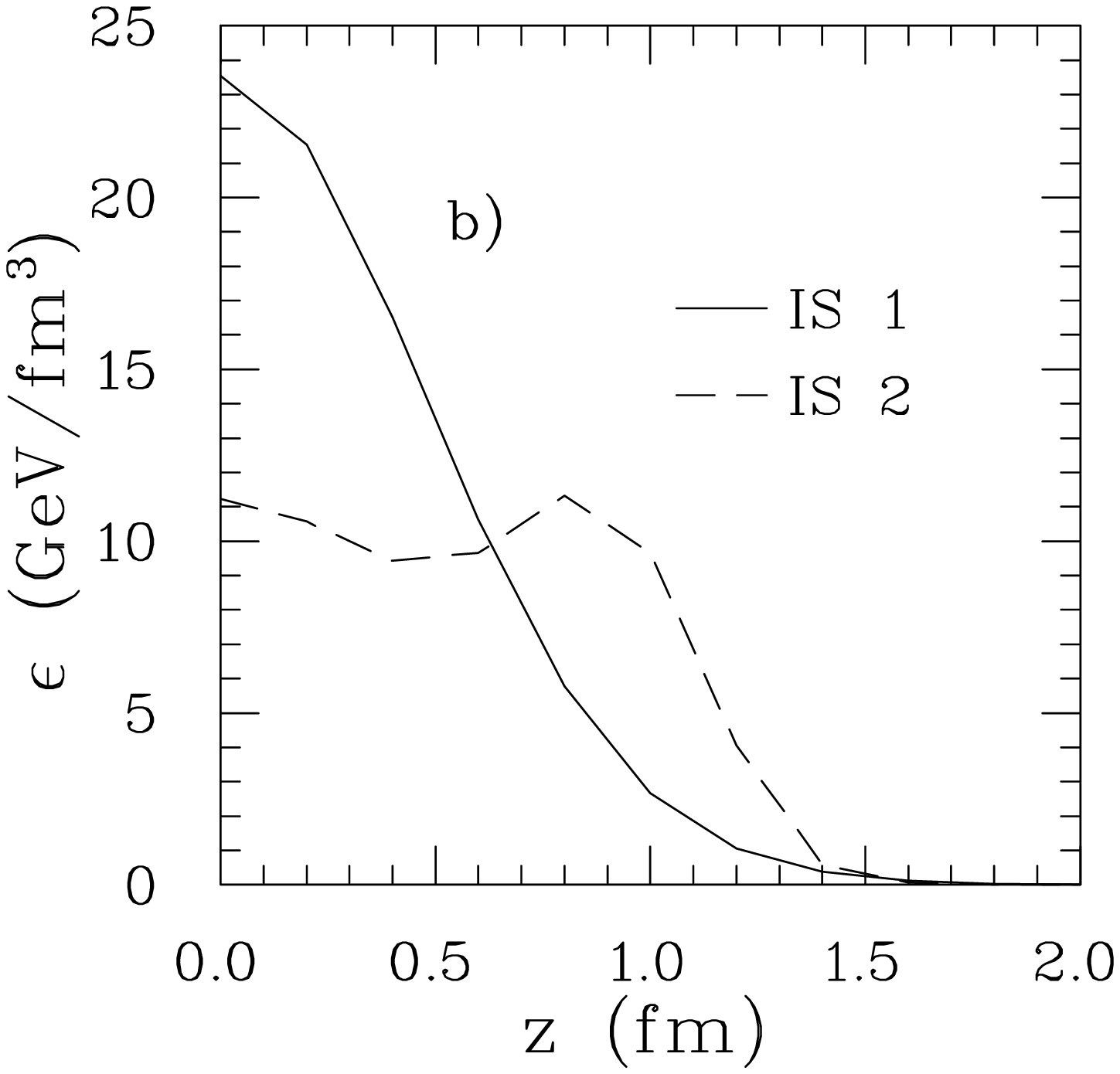}
        \hfill
  \end{minipage}
  \caption{Initial velocity profile (a) and energy density 
           profile (b) for EoS~D.}
  \label{initial}
\end{center}
\end{figure}

At low energies a hydrodynamical description of the whole collision,
including also the initial
compression and heating of the nuclear matter is feasible. At the SPS
energy  range nuclear transparency is already present and the physical
picture of compressing fluids does not apply. On the other hand the
main entropy production phase becomes of order 1--2 fm/$c$,
considerably less than the total time interval of secondary
interactions, $\approx$ 10 fm/$c$. It is then reasonable to consider
the production and the expansion separately. The simplest approach,
which we adopt, is to assume that the initially produced particles
quickly form a thermal system and to parametrize it in terms of
densities and fluid velocity.

We have presented our model for parametrizing the initial state in
terms of local nuclear thickness in 
refs.~\cite{Sollfrank97b,Sollfrank97c}. Our parametrization gives
energy and baryon number distributions in rapidity space, which leaves
us the freedom to change the initial spatial distributions by changing
the initial velocity profile. A velocity profile where rapidity $y$ has
a linear $z$-dependence \cite{Sollfrank97b,Sollfrank97c} leads to
spatial energy distribution which is peaked in the middle of the system
(fig.~\ref{initial}, solid line). To make the energy distribution less
peaked we change the velocity profile to have a smaller slope
${\rm d} v/{\rm d} z$ at small $z$ and deeper slope at larger $z$ 
(fig.~\ref{initial}, dashed line). This way we shift initial thermal
energy to initial kinetic energy. The volume of the fireball inside
the freeze-out surface stays unchanged which ensures that the states
can not evolve to each other but are genuinely different. To reproduce
the hadronic data the use of the new velocity profile requires less
stopping compared to the former linear profile in $y$.

\begin{table}
\begin{center}
\begin{tabular}{|l||c|c|c||c|c|c||}
\hline
       & \multicolumn{3}{c||}{IS 1} & \multicolumn{3}{c||}{IS 2} \\ 
                                                                 \hline
             & EoS A & EoS~D    & EoS H   & EoS A  & EoS D & EoS H \\
                                                                 \hline
$\overline{\varepsilon} $ (GeV/fm$^3$)
             & 10.2  &  10.3    &  10.3   &  5.0   &  5.3  & 5.3  \\
$\overline{\rho}_{B} $ (fm$^{-3}$)
             &  1.1  &   1.0    &  1.0    &  0.59  &  0.59 & 0.59 \\
$\overline{T} (z=0)$ (MeV)
             &  255  &   258    &  234    &  214   &  220  & 213  \\
\hline
\end{tabular} \vspace*{2ex}
 \caption{Average initial densities and temperature in the central
          region of the fireball for each  EoS and initial state. 
          Average values of energy and baryon number density are
          calculated in a region where flow rapidity is 
          $-0.5 < y <0.5$. Average temperature is calculated at $z=0$.}
 \label{tab1}
\end{center}
\end{table}

From now on we refer to these different initial states as IS~1 and IS~2
meaning a state with a peaked energy distribution and a state with 
a wide energy distribution respectively (see fig.~\ref{initial}).
The velocity profiles are the same for all EoSs but the energy
density and baryon number distributions vary slightly to fit the data.
However the shapes of the distributions are as depicted in 
fig.~\ref{initial} for all the EoSs. Resulting average densities and
temperature in the central region of the fireball are listed in 
table~\ref{tab1}.

\section{Emission rates}

The spectra of dileptons and photons are calculated using thermal
emission rates for the different phases. In QGP we use the lowest order
rates both for the lepton pairs~\cite{Cleymans87} and 
photons~\cite{Traxler95}. For photons the screening of the quark mass
singularity through the resummation of hard thermal loops in the quark 
propagator is included \cite{Traxler95}. For lepton pairs with masses
greater than the temperature, next to leading order corrections are not
very significant. At small masses they become
large~\cite{Braaten90,Altherr92} but the rate is then much below those
from other sources.

In the hadron gas many different processes contribute. These have been
considered by Gale and Lichard \cite{Gale94} for interactions among
vector and pseudoscalar mesons and the authors have provided us with
their numerical result. Since the vector mesons are included in the
form factors in the cross sections, e.g.\ the $\rho$ pole in the
$\pi^+\pi^-$ annihilation to lepton pairs, their decays during the
thermal stage are not counted separately. At intermediate masses the
process $\pi a_1 \to l\bar{l}$ is also important. Gale and 
Li~\cite{Gale98} calculated the production rates in the hadron gas
including this process. The rates due to $\pi a_1 \to l\bar{l}$
process were calculated in three different ways. Those calculated
assuming that the $\pi a_1$ electromagnetic form factor is represented
by the $\rho (770)$ only give the best fit to the data~\cite{Gale98}.
Therefore we use those rates, provided kindly by C.~Gale, at masses
above 1 GeV.
The use of the largest production rates presented
in~\cite{Gale98} would enhance the thermal dilepton production in our
calculations at largest by a factor 1.5 at 1.5 GeV mass. Below 1.2 GeV
and above 2.5 GeV masses the difference is negligible.

The photon emission in the hadron gas was calculated in
ref.~\cite{Kapusta91} using a pseudoscalar-vector Lagrangian to
describe the interactions among the mesons. A useful parametrization of
the rates is given in ref.~\cite{Nadeau92}. An important contribution
from the $\pi-\rho$ channel involving the formation and decay of the
$a_1$ axial vector meson \cite{Xiong92} is also included. As for the
lepton pairs, rates from processes involving baryons are not included.

Given the local emission rates we can calculate the total emission in a
nucleus-nucleus collision by folding the rates with the flow. E.g.~for
lepton pairs we have
\begin{eqnarray}
  \frac{dN^{l^+l^-}}{d^4p} 
    = \int d^4x \left\{ w(\epsilon,\rho_B) 
                       \frac{dR^{QGP}}{d^4p}(p\cdot u,T,\mu_B)\right.
    & & \nonumber \\
          \left.      +[1-w(\epsilon,\rho_B)]
                       \frac{dR^{HG}}{d^4p}(p\cdot u,T) \right\} & &
\end{eqnarray}
where $u\cdot p$ is the energy of the pair in the local rest frame of
the fluid element with four-velocity $u$. The emission rate per unit
volume and time is $dR/d^4p$ and $w(\epsilon,\rho_B)$ is the fraction
of plasma phase at space-time point $x$. In the QGP $w=1$, in the
hadron gas $w=0$ and in the mixed phase $0 < w < 1$.

\section{Results}

	\subsection{Hadron spectra}
	  \label{hadronsection}

Hadron spectra have already been studied in more detail
by several groups 
\cite{Schlei97,Cleymans97,Sollfrank97c,Ornik96,Dumitru98}
within the local hydrodynamical model.
In ref.~\cite{Sollfrank97c} we have discussed the hadron spectra for
$B + A$ collisions measured at CERN. As new features we consider here 
the effects of freeze-out and phase transition temperatures on the
spectra of negative particles and net protons. We also study whether
different initial conditions may lead to an adequate description of
hadronic spectra. Particle spectra are calculated using the
prescription of Cooper and Frye~\cite{Cooper74} including the same
hadrons as in the construction of the equation of state. In Cooper's
and Frye's prescription an important notion is the freeze-out surface
separating the regions where particles behave as hadron gas and as
free particles. Freeze-out takes place where the mean free path of
particles is of the same order than the size of the fireball. For the
collisions of light nuclei this criterion leads to a freeze-out
temperature of $T_f \approx 140$ MeV~\cite{Goity89}. However, it has
been suggested recently that a freeze-out temperature of 
$T_f \approx 120-130$ MeV would be more appropriate for a Pb+Pb 
collision~\cite{Kampfer96,Wiedemann98,Appelshauser98}.
Instead of having the freeze-out
to take place at constant temperature, we define the freeze-out on a
space-time surface of constant energy density. As a freeze-out energy
density we use $\epsilon_f = 0.15$ GeV/fm$^3$ and
$\epsilon_f = 0.069$ GeV/fm$^3$ which result in an average freeze-out
temperature of $T_f \approx 140$ MeV and $T_f \approx 120$ MeV,
respectively.

\begin{figure}
\begin{center}
  \begin{minipage}[t]{\size}
        \epsfxsize \size \epsfbox{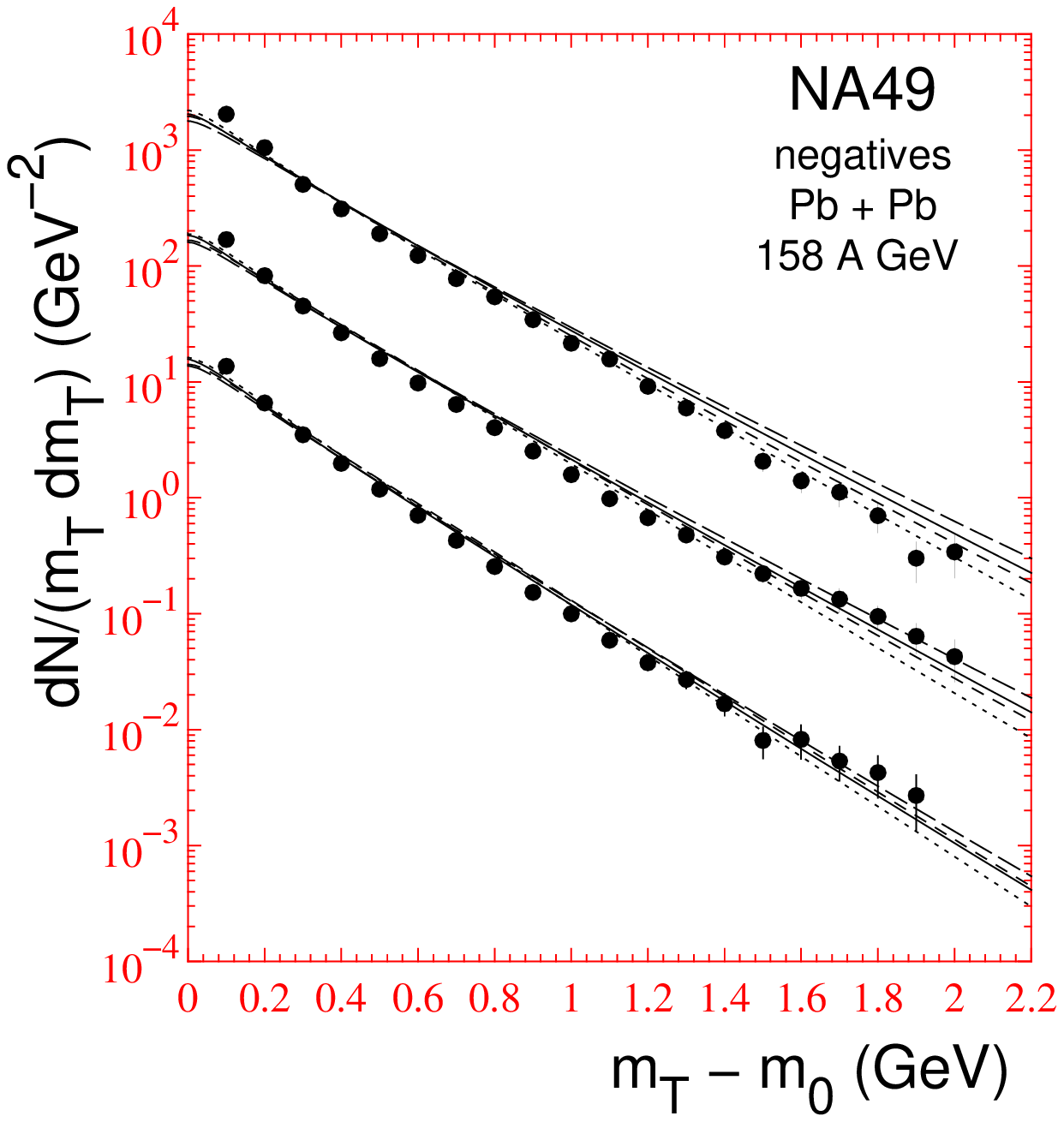}
        \hfill
  \end{minipage}
  \hfill
  \begin{minipage}[t]{\size}
        \epsfxsize \size \epsfbox{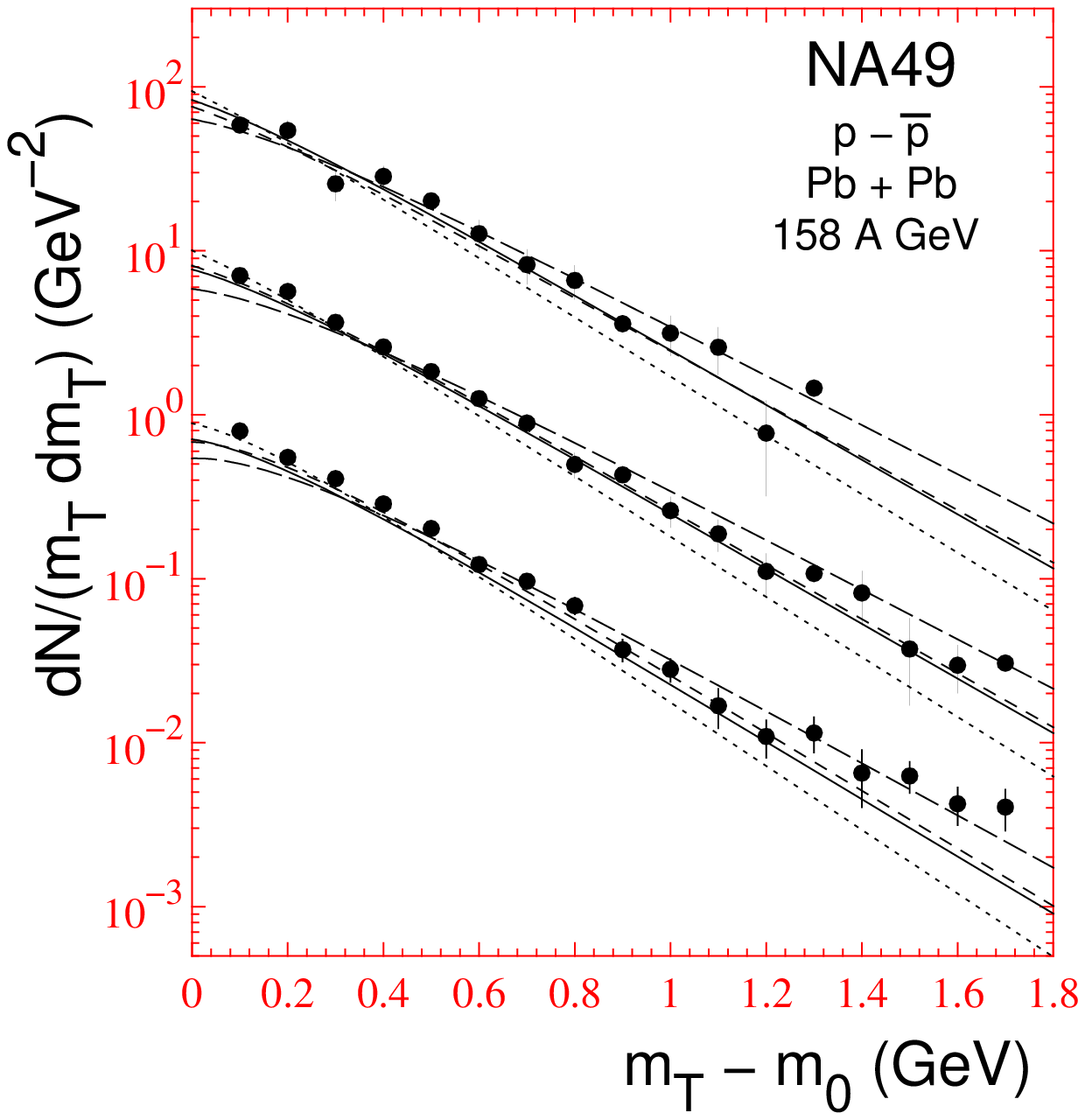}
        \hfill
  \end{minipage}\\

  \vspace*{-1.0truecm}
  \begin{minipage}[t]{\size}
        \epsfxsize \size \epsfbox{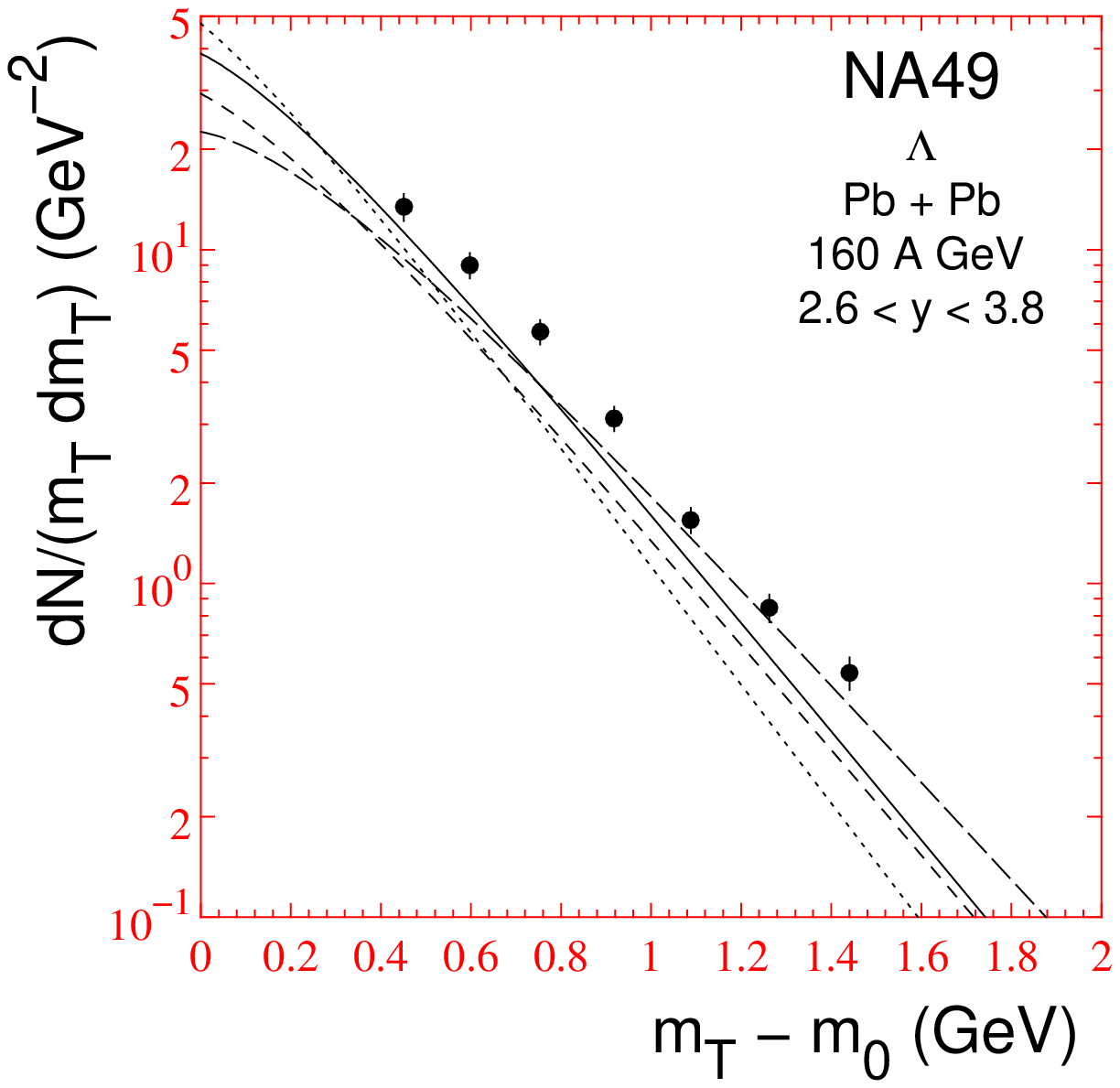}
        \hfill
  \end{minipage}
  \hfill
  \begin{minipage}[t]{\size}
        \epsfxsize \size \epsfbox{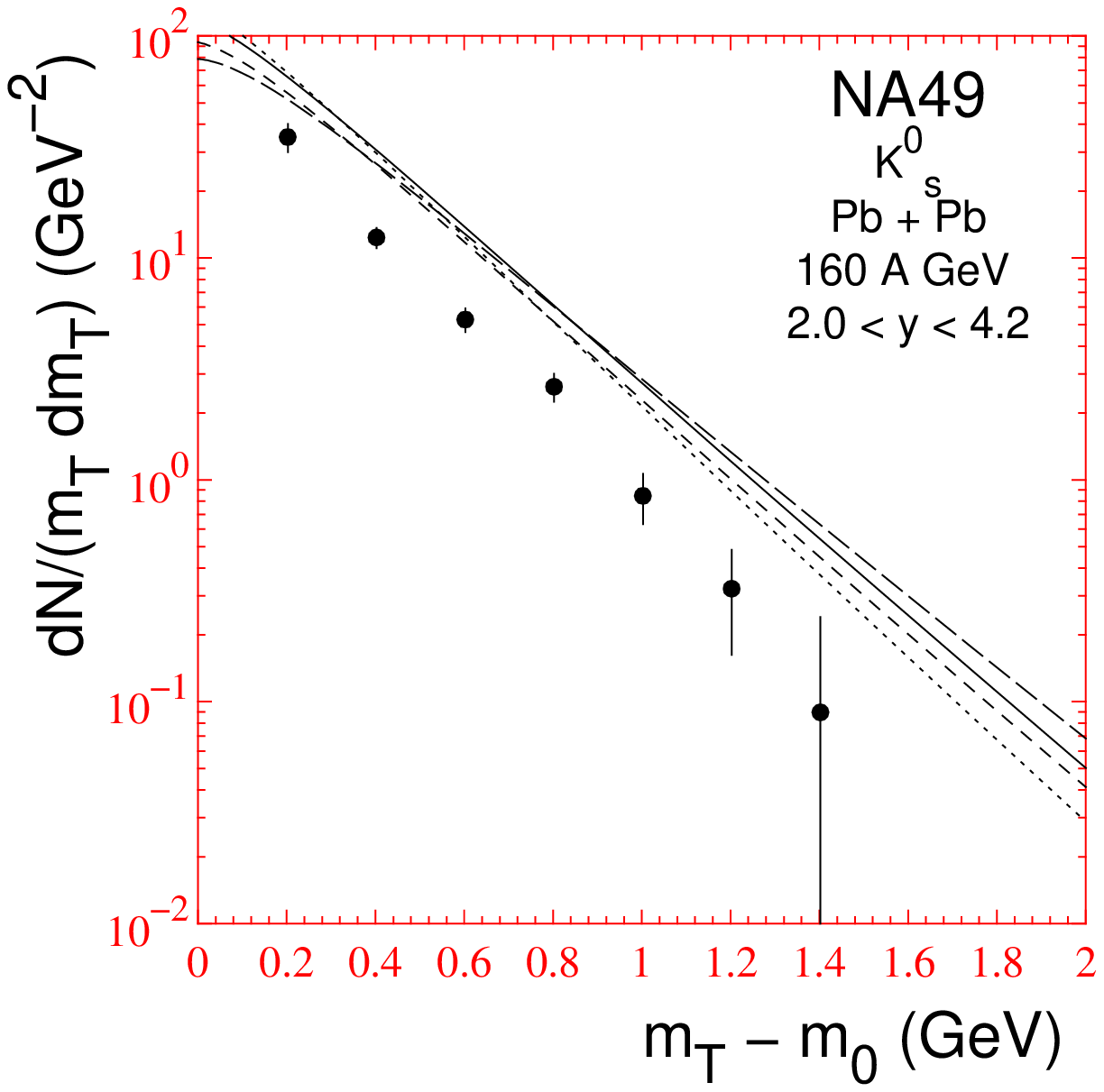}
        \hfill
  \end{minipage}
  \caption{Transverse momentum distributions of negative particles,
           net protons, lambdas and neutral kaons for different EoSs and 
           freeze-out temperatures.
           Solid line is EoS~D with $T_f \approx 140$ MeV,
           long dashed line EoS~D with $T_f \approx 120$ MeV,
           dotted line EoS~A with $T_f \approx 140$ MeV and
           short dashed line EoS~A with $T_f \approx 120$ MeV.
           Initial state is IS~1.
           Data are from the NA49 collaboration \cite{Jones96,Jacobs97}.
           Negative particle data are for rapidity
           intervals of width 0.5 and center at (top to bottom)
           3.4, 3.9, 4.4.
           Net proton data are for rapidity intervals of
           width 0.5 and center at (top to bottom) 2.9, 3.4, 3.9.
           In both cases the data sets are successively scaled down
           by $10^{-n}$, $n$=0,1,2.}
  \label{hadron2}
\end{center}
\end{figure}

The effect of the freeze-out temperature on transverse momentum spectra can be
seen in fig.~\ref{hadron2} where we show the $p_T$ spectra of negative
particles, net protons, lambdas and neutral kaons using decoupling 
temperatures
$T_{f}\approx 140$ MeV and $T_{f}\approx 120$ MeV. The spectra are
obtained using EoSs A and D and initial state IS~1. The effect of the
freeze-out temperature is similar when initial state IS~2 is used.
Lower freeze-out temperature leads to stronger flow but for light 
particles this effect is counterbalanced by the decrease in
temperature. Therefore the spectra of negative particles, mostly pions,
are almost unaffected by the change in freeze-out temperature. The
difference in slope constant $T$, defined in the same way as for the
data in~\cite{Jones96}, is about 10 MeV for all EoSs and both initial
conditions. On the other hand heavier protons gain more from the
increased flow velocity than they lose with decreasing temperature.
The lower freeze-out temperature leads to 30--45 MeV larger slope
constants than $T_f \approx 140$ MeV (see table~\ref{tab2}). Comparing
with data the 120 MeV freeze-out temperature produces better fits than
$T_f \approx 140$ MeV when EoS~A is used. However, EoS~D is stiffer
than EoS~A, which leads to larger flow velocity. Therefore the spectra
obtained using EoS~D and $T_f \approx 140$ MeV are very close to those
obtained using EoS~A and $T_f \approx 120$ MeV whereas the combination
of EoS~D and $T_f \approx 120$ MeV produces slightly too large flow.

The particle abundances in heavy ion collisions and thus the 
normalization of the spectra are an interesting field of research
of its own. The thermal model applied to particle ratios
and abundances gives usually reasonable results but much higher
temperatures of order 160--180 MeV \cite{BraunMunzinger96,Becattini98}. 
The difference to the 120--140 MeV
freeze-out temperature used here may be explained by a separation of
the chemical freeze-out where particle ratios are frozen from the
kinetic freeze-out where the shape of the spectra is determined.
For technical reasons it is very hard to separate these decoupling scales
in a hydrodynamical simulation. Therefore we assume chemical
equilibrium down to the kinetic freeze-out. The price we pay is 
that the normalizations of other particles than pions and nucleons
may come out wrong. As an example we show the $\Lambda$ and K$^0_s$ 
$p_T$-spectra in fig.~\ref{hadron2} for the sake of comparison. 
While the slope comes out approximately right, the normalization is
off as been expected. 
The slopes of the kaon and lambda spectra depend on the EoS and
freeze-out energy density in the same way than for pion and net
proton $p_T$ spectra: the combination of EoS~D and $T_f \approx 140$ MeV
on one hand and EoS~A and $T_f \approx 120$ MeV on the other produces
slopes which are closest to the experimental ones.

\begin{table}
\begin{center}
\begin{tabular}{|l|c||c|c|c||}
\hline
  \multicolumn{2}{|c||}{}     & $y=2.9$ & $y=3.4$ & $y=3.9$ \\ 
                                                          \hline \hline
  \multicolumn{2}{|r||}{NA 49~\cite{Jones96}} 
                          & $290\pm 20$ & $273\pm 8$ & $262\pm 8$ \\  
                                                          \hline \hline
      & $T_f\approx 120 $ MeV & 315     & 318     & 310     \\  
                                                            \cline{2-5}
 \raisebox{2.0ex}[0pt]{EoS D} & $T_f\approx 140 $ MeV
                              & 273     & 275     & 268     \\  \hline
      & $T_f\approx 120 $ MeV & 280     & 277     & 274     \\  
                                                            \cline{2-5}
 \raisebox{2.0ex}[0pt]{EoS A} & $T_f\approx 140 $ MeV
                              & 247     & 243     & 239     \\  \hline
\end{tabular} \vspace*{2ex}
  \caption{The values of the inverse slope constant $T$ in MeV for net
           proton $p_T$ spectra at various rapidities obtained using
           EoS~D and A, initial state IS~1 and freeze-out temperatures 
           $T_f\approx 140$ and 120 MeV. The accuracy of our fits 
           is about 5 MeV.}
  \label{tab2}
\end{center}
\end{table}

\begin{figure}
\begin{center}
  \begin{minipage}[t]{\size}
        \epsfxsize \size \epsfbox{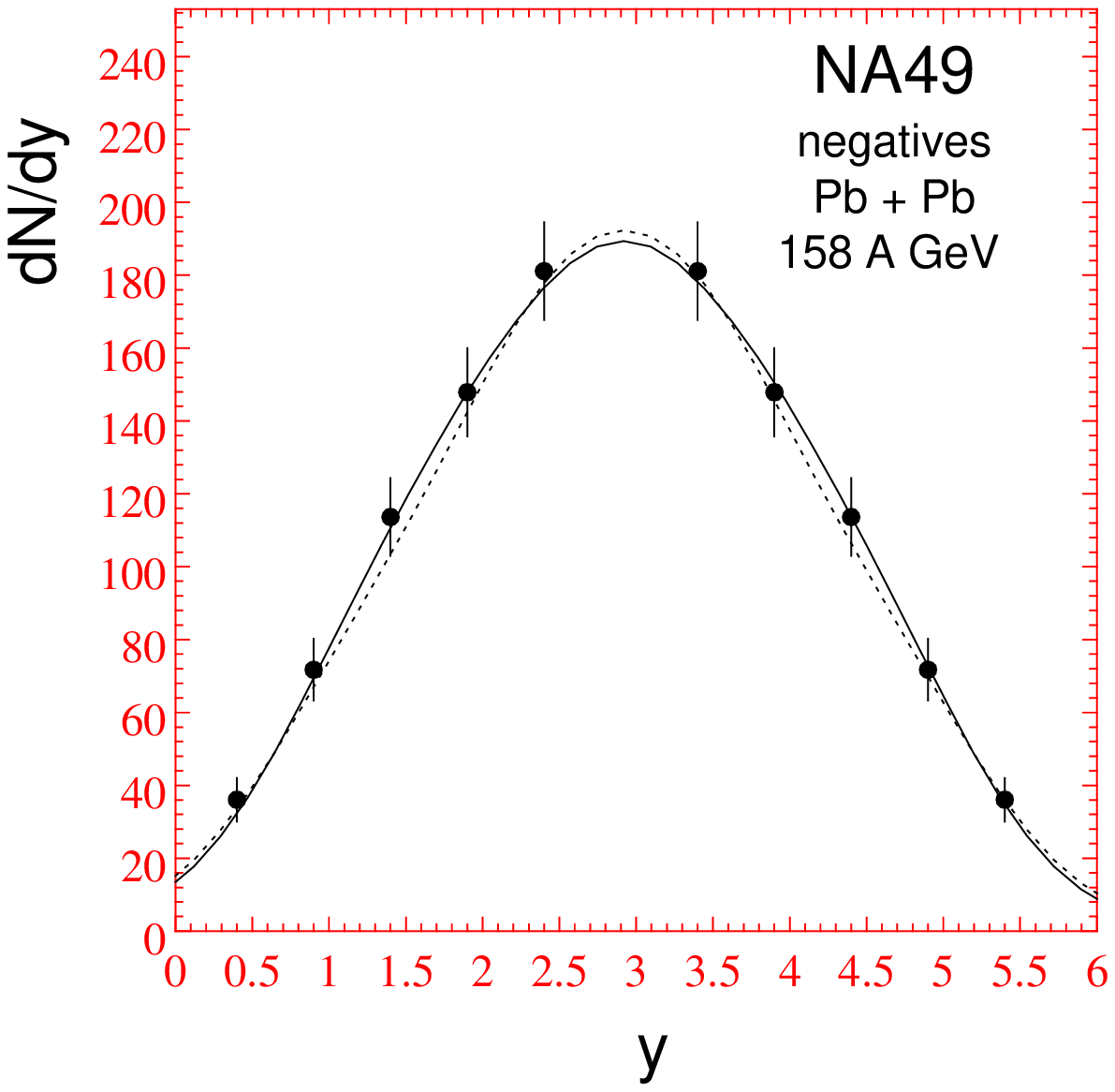}
        \hfill
  \end{minipage}
  \hfill
  \begin{minipage}[t]{\size}
        \epsfxsize \size \epsfbox{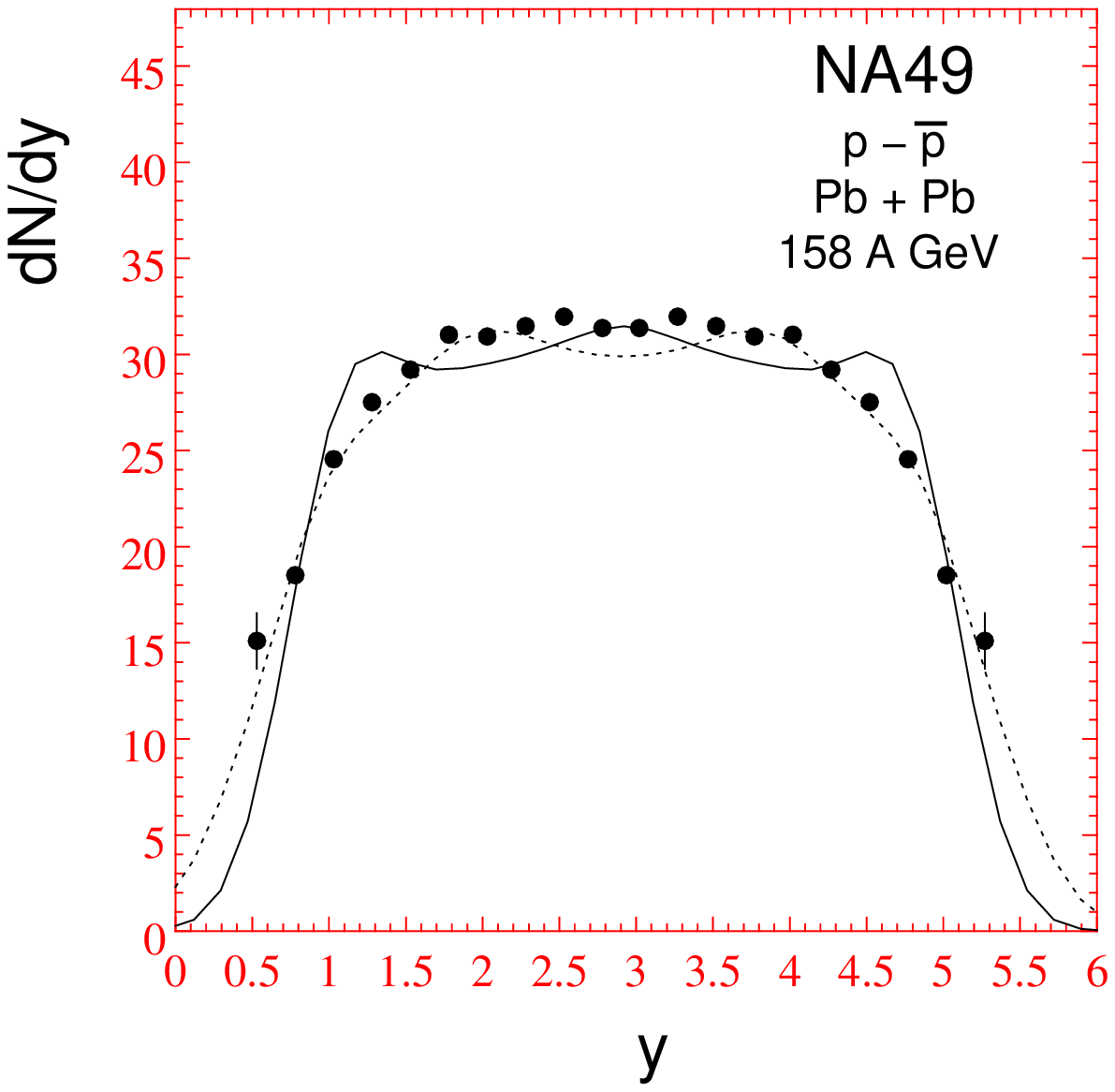}
        \hfill
  \end{minipage}\\

  \vspace*{-1.0truecm}
  \begin{minipage}[t]{\size}
        \epsfxsize \size \epsfbox{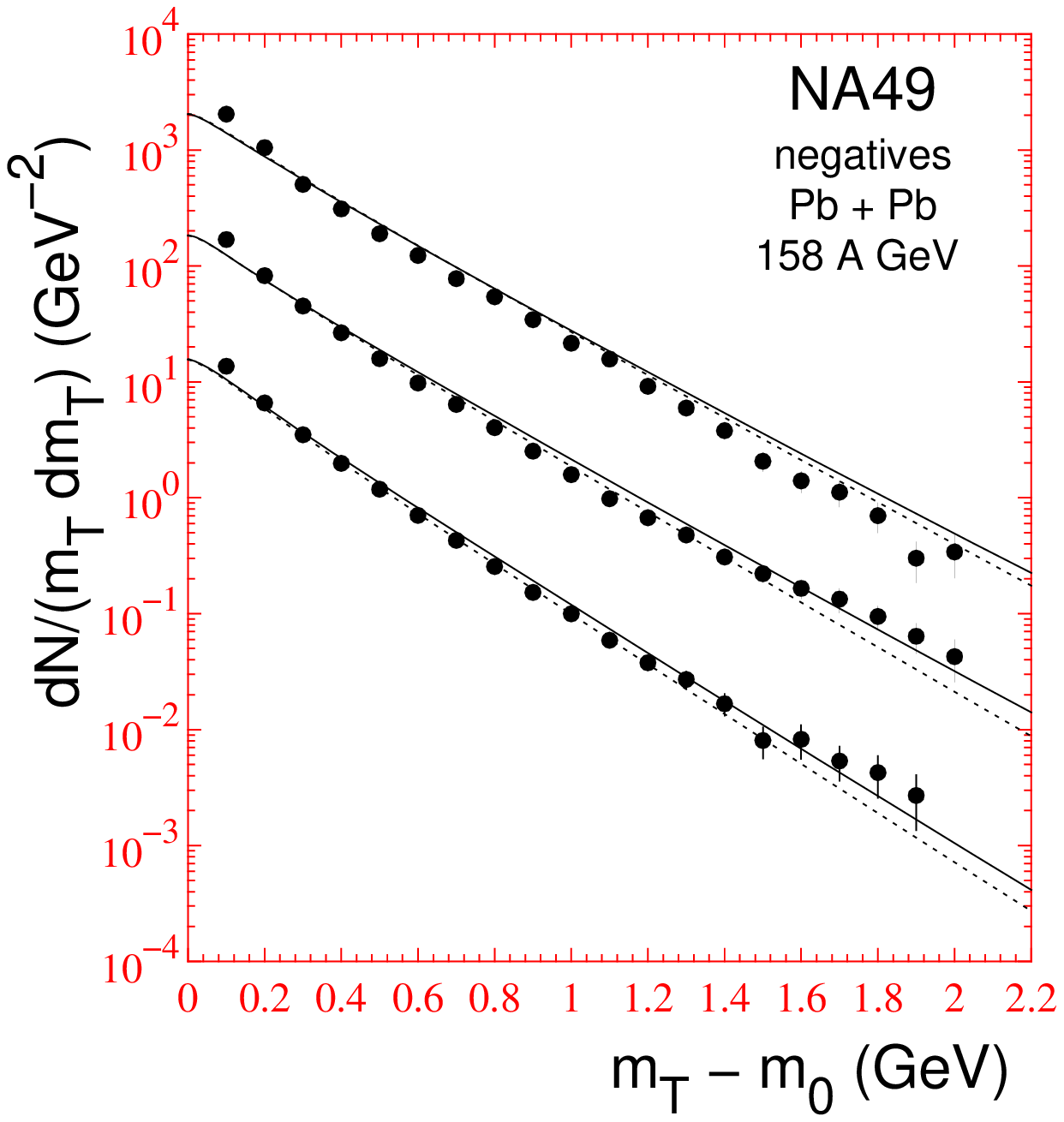}
        \hfill
  \end{minipage}
  \hfill
  \begin{minipage}[t]{\size}
        \epsfxsize \size \epsfbox{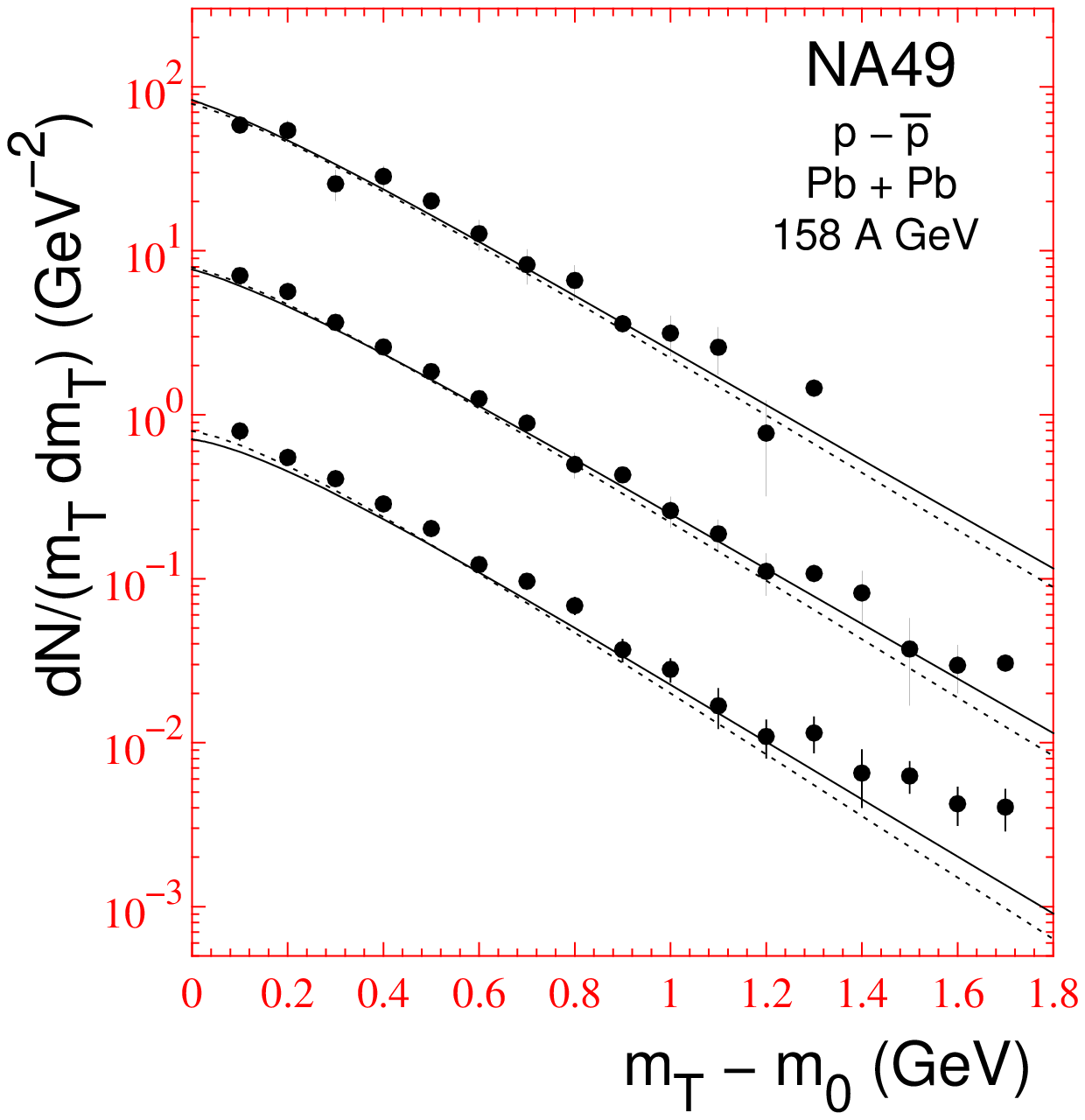}
        \hfill
  \end{minipage}
  \caption{Rapidity and transverse momentum distributions of negative 
           particles and net protons for EoS~D. Freeze-out temperature
           is $T_f \approx 140$ MeV. Solid line corresponds to initial
           state IS~1 and dotted line to IS~2. Data are from the NA49
           collaboration \cite{Jones96,Jacobs97}. Transverse momentum
           spectra are presented like in fig.~\ref{hadron2}.}
  \label{hadron1}
\end{center}
\end{figure}

As an example of the effect of the initial conditions to the spectra
we show in fig.~\ref{hadron1} the resulting rapidity and $p_T$ spectra 
of negative particles and net protons for initial states IS~1 and IS~2.
In both cases the equation of state is EoS~D and freeze-out temperature
$T_f \approx 140$ MeV. For the negative particles the calculations are in
good agreement with the data, which tells that two different initial
conditions can result in almost similar flow of energy density across
the freeze-out surface. However, the hydrodynamical evolution does not
smooth away all differences of the initial conditions since the
rapidity spectrum of net protons still shows some signs of different
initial stopping of baryons and flow in these two cases. As one may
expect the larger initial energy density and thus greater pressure
leads to stronger transverse flow, but the effect is small. The
difference in slope parameter $T$ is 10--15 MeV for net protons and
about 10 MeV for negative particles.

Also when EoSs A and H are used, the spectra obtained using initial
states IS~1 and IS~2 are very close to each other. The differences in
the slope parameter $T$ are similar to those described in the case of
EoS~D. When EoS~H without phase transition to quark-gluon plasma is
used, the resulting spectra differ only slightly from those obtained
using EoS~D. The slope parameters of the net proton $p_T$ spectra are
10--15 MeV larger for EoS~H than for EoS~D. For negative particles the
differences in the slope parameters are below 5 MeV. The initial state
and hadron spectra are almost identical for EoS~D and H because in
both cases the hydrodynamical evolution is dominated by the hadronic
part of the EoS. Plasma and mixed phase last too short a time and
occupy too small a volume to affect the flow considerably.

Independently of initial state the best fit is obtained combining
EoS~A with $T_f \approx 120$ MeV and either EoS~D or EoS~H with 
$T_f \approx 140$ MeV. We conclude that the initial state of a
hydrodynamic description is not unambiguous even for a given EoS.
Different initial conditions can lead to an acceptable description of
the data if the densities and the velocity profile are correlated
carefully.

	\subsection{Photons}

\begin{figure}
\begin{center}
  \begin{minipage}[t]{\sizetwo}
        \epsfxsize \sizetwo \epsfbox{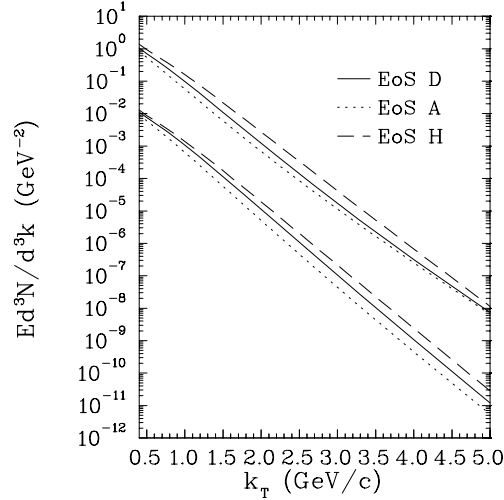}
        \hfill
  \end{minipage}
  \caption{Single photon $k_T$ spectra for three different equations
           of state and initial states IS~1 (upper three lines) and
           IS~2 (lower three lines). Photon spectra for IS~2 are
           scaled down by 10$^2$. Freeze-out temperature is 
           $T_f \approx 140$ MeV for both EoSs D and H, and 
           $T_f \approx 120$ MeV for EoS~A.}
  \label{photfig}
\end{center}
\end{figure}

Unlike hadrons, photons are emitted during the whole lifetime of the
fireball and may thus carry information from the hot and dense stage
of the collision. Since we found in the previous section that using
different freeze-out temperature for different EoSs gave the best fit
to the data, we use from now on EoS~A with $T_f \approx 120$ MeV and
EoSs D and H with  $T_f \approx 140$ MeV. In fig.~\ref{photfig} we show
direct photon $k_T$ spectra for three different EoSs and for initial
states IS~1 (upper three lines) and IS~2 (lower three lines scaled down
by $10^2$). In the case of EoS~A and IS~1 the contribution from the
plasma phase dominates at $k_T > 2.5$ GeV/$c$ resulting in a
distinctive concave shape of the spectrum. When EoS~D with higher phase
transition temperature is used together with IS~1, the contribution from the
hadron gas dominates up to $k_T = 3.5$ GeV/$c$ and the shape of the
spectra turns concave at larger transverse momenta. Our results are 
very similar to those presented in~\cite{Hirano97}, especially for EoS~A,
whereas the photon yield presented in~\cite{Cleymans97,Sarkar98}
is in the case of a phase transition much smaller
at large values of $k_T$ than our result.
This is due to the smaller initial temperature and the use of boost invariant
hydrodynamics in~\cite{Cleymans97,Sarkar98}, 
which leads to weaker transverse flow than our approach.

For the initial state IS~2 the photon emission at $k_T < 2$ GeV/$c$
is very close to that obtained using IS~1, but the difference
increases rapidly with $k_T$. At larger values of $k_T$ the emission
is dominated by the early, hot stage of the evolution. Therefore
decreasing initial temperature cuts the emission accordingly. This can
be seen very clearly when EoS~A is used. Even if the lifetime of the
plasma phase is almost equal in both cases, the drop in initial
temperature causes one order of magnitude change in photon yield when
$k_T > 4$ GeV/$c$ and the clearly concave shape has almost vanished.
For EoS~D and H the effect is smaller. At high values of transverse
momenta the photon yield is changed by a factor of 6 or 4 respectively.
Thus the effect of initial densities on photon emission at high values
of $k_T$ depends on the EoS.  

Because the photon emission rate in the hadron gas is larger than in
the plasma, the photon yield is the larger the higher the phase
transition temperature is. However, due to lower freeze-out
temperature used with EoS~A, the differences between the total yields
are smaller than expected from the dependence on the phase transition
temperature. When IS1 is used EoS~D
increases the photon yield by a factor of 1.2--1.8 and EoS~H by a factor
of 1.6--4 compared to EoS~A.
As mentioned, the use of IS~2
cuts high-$k_T$ photon production especially for EoS~A and therefore
the differences between EoSs are larger. When IS~2 is used, 
the photon spectra differ by factors of 1.2--2.3 (EoS D compared to EoS A)
or 1.4--5.5 (EoS H compared to EoS A),
the difference increasing with $k_T$. For IS~1 the difference between
EoSs is largest at $1.5<k_T<3$ GeV/$c$.

	\subsection{Lepton pairs}

	\subsubsection{Mass spectrum}

We compare our results with the data measured by the CERES 
collaboration~\cite{cer,Ravinovich98}.
These measurements are made for Pb+Au collision at 158 $A$ GeV energy,
which should be comparable with our results for Pb+Pb collision at
the same energy.

In the CERES measurements the dielectron background from the decays of
final mesons is not subtracted and the measured dilepton yield consists
of two parts. Instead of using the background estimated by the CERES
collaboration~\cite{cer}, we calculate this background from our
hydrodynamical simulation (see ref.~\cite{Sollfrank97a}) using thermal
yields for decaying mesons. However, our model assumes chemical
equilibrium but the observed $\phi/h^-$ ratio~\cite{Friese97} is
equivalent with the chemical equilibrium value only around the
temperature $T \approx 120$ MeV \cite{Becattini98}.
Since thermal models~\cite{BraunMunzinger96,Becattini98} show that the system
is not in complete chemical equilibrium at the time of kinetic freeze-out
we do not take this as an indication of a lower freeze-out temperature
but as an indication of strangeness non-equilibrium instead (see
section~\ref{hadronsection}). To achieve
consistency with the data we suppress the $\phi$-yield by a factor of
0.6 when we use the freeze-out temperature $T_f \approx 140$ MeV. 
Like for photons we show here results obtained using EoS~A with
$T_f \approx 120$ MeV, and EoSs D and H with $T_f \approx 140$ MeV.
We have also checked that the number of $\eta$-mesons obtained in
our model is consistent with the experimental results~\cite{Geurts98}.

\begin{figure}
\begin{center}
   \begin{minipage}[t]{\size}
         \epsfxsize \size \epsfbox{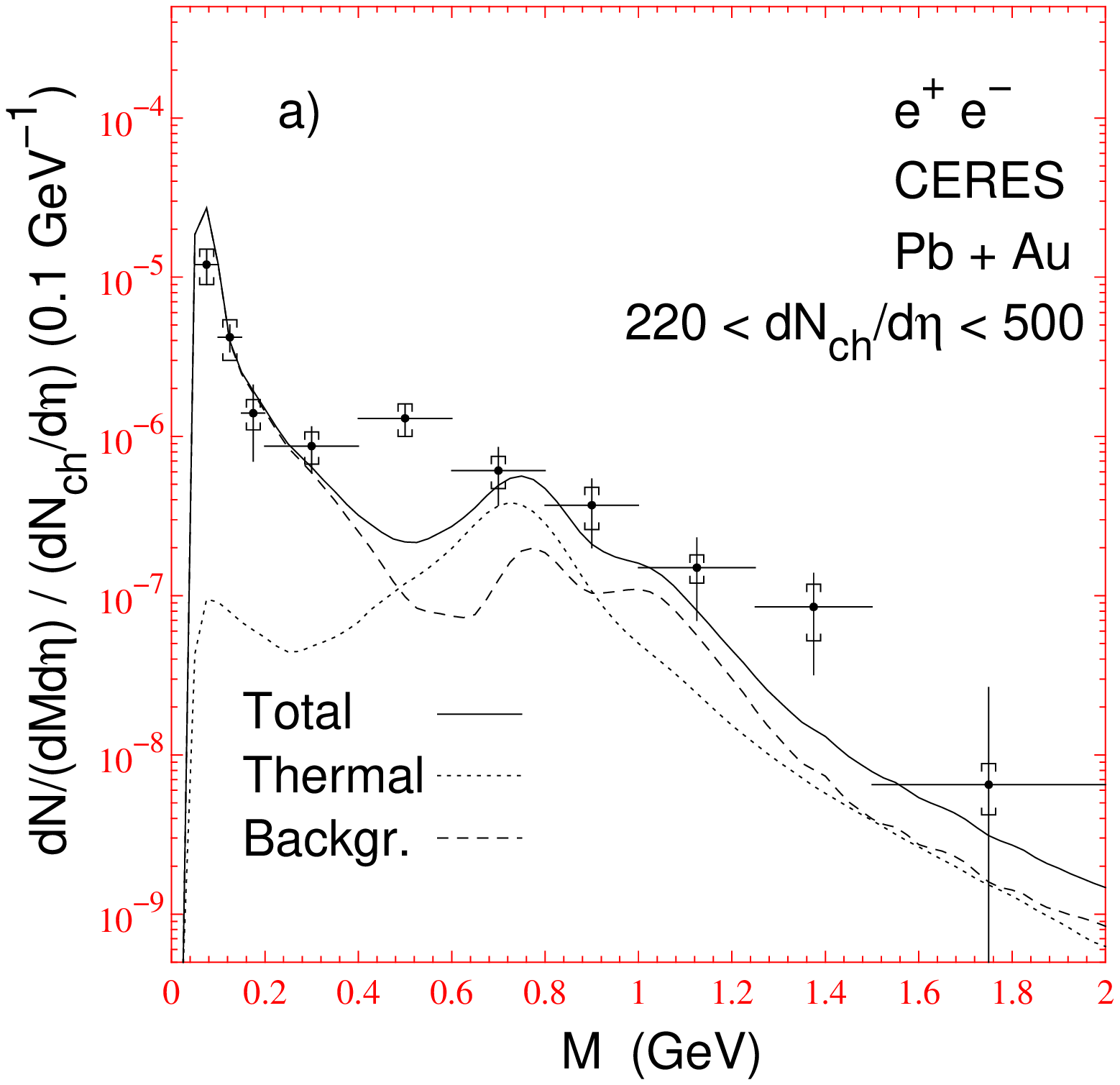}
         \hfill
   \end{minipage}
    \hfill
   \begin{minipage}[t]{\size}
         \epsfxsize \size \epsfbox{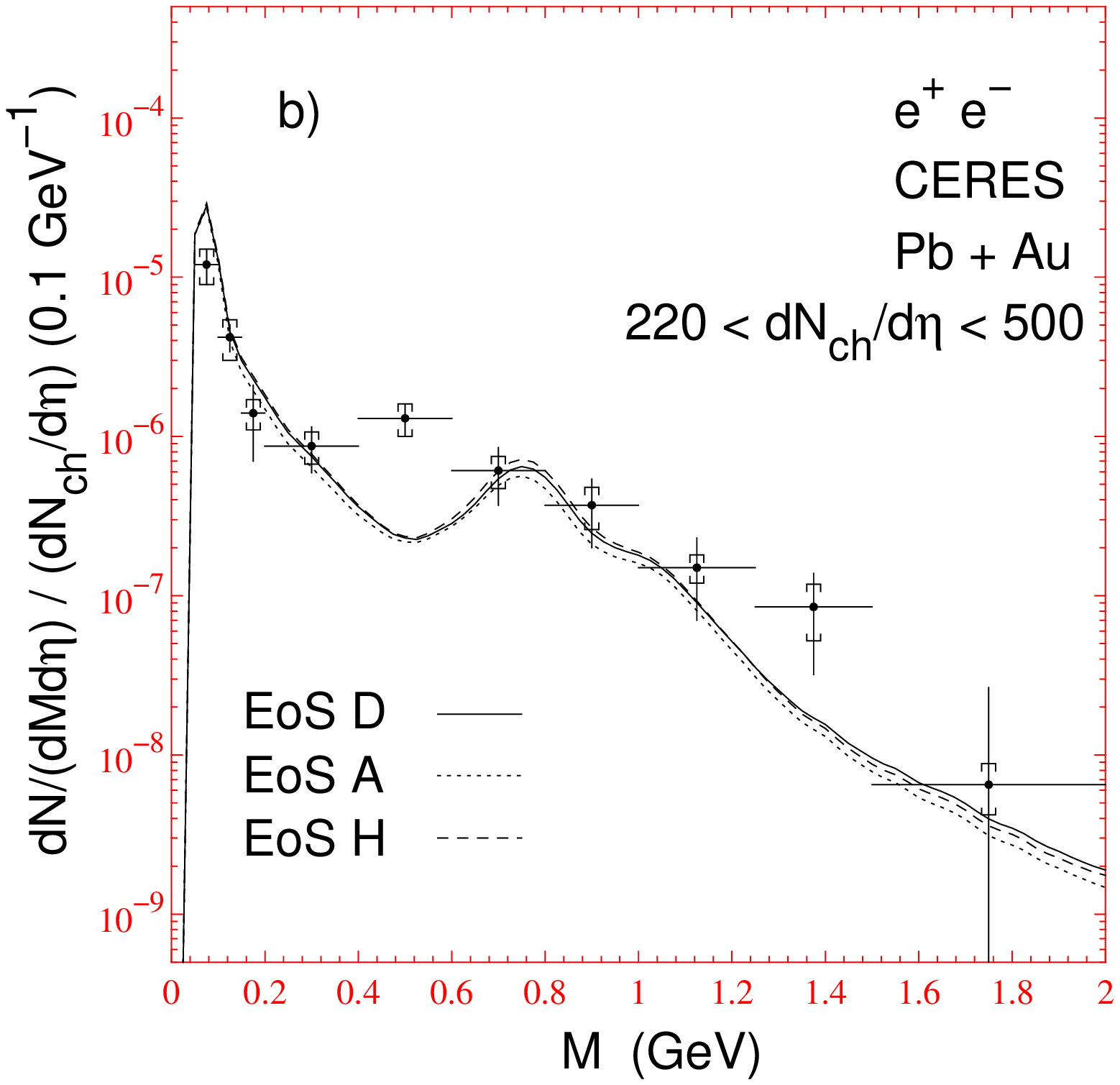}
         \hfill
   \end{minipage}
   \caption{Calculated electron pair spectrum compared to CERES 
            data~\cite{Ravinovich98}.
            (a) The emission from the fireball, background and the
            total spectra for EoS~A and IS~1. (b) The total spectrum
            using three different EoSs and IS~1. The kinematic cuts
            and the detector resolution of the CERES experiment are
            incorporated.}
   \label{ceres}
\end{center}
\end{figure}

In fig.~\ref{ceres}a we show the background, the yield from thermal
emission and the combined total mass spectrum folded with the CERES
cuts and resolution. The thermal emission is calculated with EoS~A and
IS~1. Note that we compare our results with data
from~\cite{Ravinovich98} for which the charged particle multiplicity
is $220 < dN_{ch}/d\eta < 500$. Our calculation is tuned to reproduce
the results of the NA49 experiment which uses a centrality trigger
different from the trigger of the CERES experiment. Hence we get an
average multiplicity of $\langle dN_{ch}/d\eta\rangle \approx 330$.
The CERES collaboration has reported dependence on the multiplicity
both in the shape and the magnitude of the spectrum scaled with
multiplicity. Therefore we think our results should be compared with
the high multiplicity data set rather than with
$\langle dN_{ch}/d\eta\rangle = 220$ data presented in~\cite{cer}. 

Our background overshoots the data at the mass of 75 MeV. This is due
to our pion spectra being too flat at low values of transverse momenta
as seen in fig.~\ref{hadron2} for negative particles, mostly pions.
Because our overall normalization is correct, we get too many pions in
the CERES acceptance region.

Our calculation fails to reproduce the observed excess of electron 
pairs~\cite{Ravinovich98} around $M\approx 500$ MeV as also the
other hydrodynamical models do without in-medium changes of particle
properties \cite{Srivastava96,Hung97,Drees96}.
At 500 MeV the
thermal emission is of the same order of magnitude than the background
but the total yield is below the data by a factor 5--6, which requires
an enhancement factor of 10 if the thermal emission is the origin of
the excess. Possible explanation of the enhancement in this mass
region have been studied extensively by other groups and are
reviewed in \cite{Wambach98}. For our studies here, it is important
that possible in-medium modifications of $\rho$-meson parameters
which might cause this enhancement should not, however, suppress
dielectron emission around free $\rho$ mass where the data are
reproduced. A study of medium modified rates is done in a later
work~\cite{Huovinen98}. 

The dilepton results are similar for all EoSs, see fig.~\ref{ceres}b. 
Despite the longest lifetime the yield obtained using EoS~A is smallest
at all values of invariant mass, but the differences are small. 
The differences are largest around the free $\rho$-meson mass and
at masses $M > 1.5$ GeV where the spectra differ by a factor 1.2--1.3.
 
At masses below 1.5 GeV the effect of initial densities is even
smaller. For EoS~H the change in total yield is smallest, below a
factor 1.1 at all values of invariant mass. For EoS~A and D the
difference is below 5\% at $M < 1$ GeV and increases slowly at larger
values of mass. At $M=2$ GeV the initial state IS~1 produces 20\%
more electron pairs than the initial state IS~2. These differences are
smaller than the present experimental errors.

\begin{figure}
\begin{center}
   \begin{minipage}[t]{\sizetwo}
         \epsfxsize \sizetwo \epsfbox{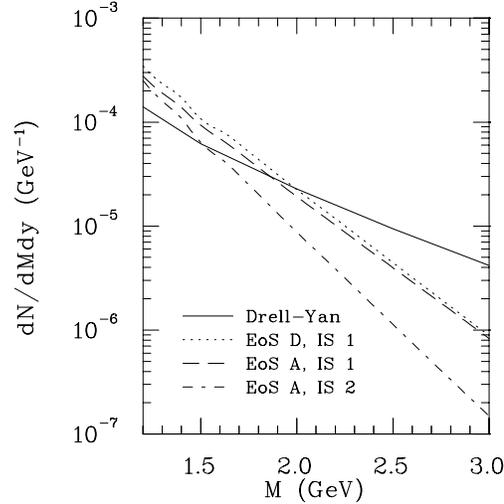}
         \hfill
   \end{minipage}
   \caption{Mass distribution of thermal electron pairs without
            kinematic cuts compared with Drell-Yan pairs calculated
            at $y_{CM}=0$.}
   \label{lepm}
\end{center}
\end{figure}

The NA50 collaboration has measured dimuon emission in the 
intermediate mass region ($1.5 < M < 2.5$ GeV)~\cite{Scomparin96},
but since the NA50 acceptance cuts are difficult to
implement~\cite{Lourenco}, we present the thermal spectra at masses
$1.2 < M < 3$ GeV without any kinematic cuts and compare them with
Drell-Yan pairs in fig.~\ref{lepm}. The Drell-Yan yield is calculated
in next to leading order using the parton distribution set CTEQ-3M
from PDFLIB~\cite{pdf}. In the analysis of NA50 the data are roughly a
factor 2 above the sum of Drell-Yan pairs and pairs from charm decays.
Depending on the equation of state and the initial condition
we found a thermal contribution of the same order as the Drell-Yan pairs
or even larger explaining semi-quantitatively the found enhancement.
A more detailed quantitative comparison with the NA50 data
in \cite{Shuryak94,Srivastava96,Li98} shows the important contribution
from secondary interactions in this mass region. 

For $M \gg T$, the emission is dominated by the early,
hot stage of the evolution. The thermal contribution is therefore
sensitive to the initial densities. This can be seen in
fig.~\ref{lepm} where the thermal emission for EoS~A is shown for
initial state IS~1 (dashed line) and IS~2 (dashed dotted line). The
dotted line depicts the spectrum with EoS~D and IS~1. When EoS~D is
used, thermal emission has qualitatively similar dependence on the
initial state with 
the yield of 3 GeV electron pairs differing by a factor 3.5.
We may thus conclude that in the intermediate mass region the
thermal lepton pair production can differentiate between different
initial conditions and that this contribution can be an important part
of the observed excess in this mass region. However, in \cite{Li98}
the missing dimuons are generated by hadronic channels only 
dominated by the $a_1$ contribution, while our calculation has
important contributions from the QGP. Therefore we consider the
question of the relevant degrees of freedom for the thermal
production of intermediate mass dimuons to be still open.

	\subsubsection{Rapidity spectrum}

\begin{figure}
\begin{center}
  \begin{minipage}[t]{\size}
        \epsfysize \size \epsfbox{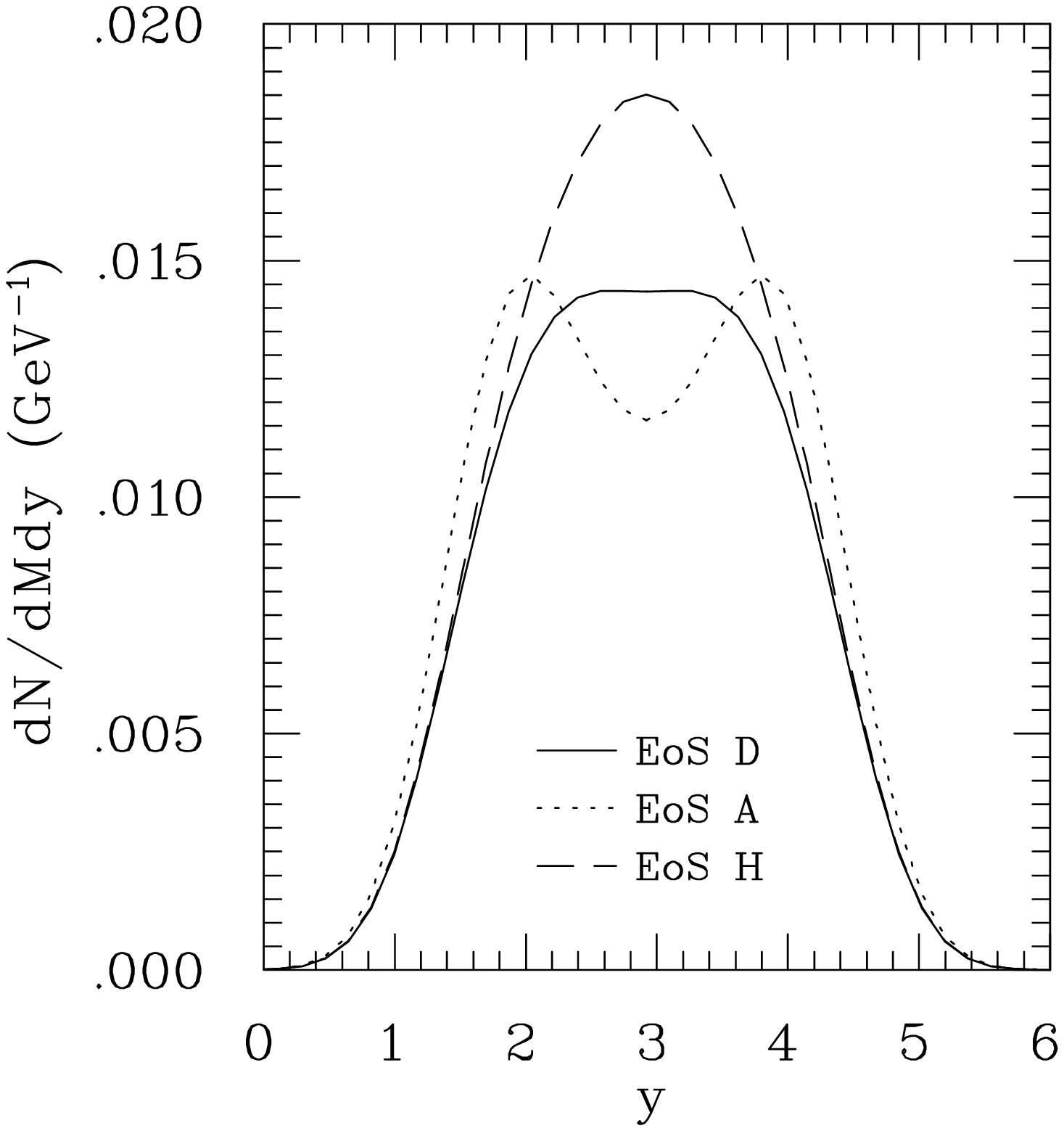}
        \hfill
  \caption{Rapidity distributions of thermal electron pairs of mass
           770 MeV. Initial state is IS~1. No kinematic cuts or
           detector resolution has been applied.}
  \label{lepy10}
  \end{minipage}
   \hfill
  \begin{minipage}[t]{\size}
        \epsfysize \size \epsfbox{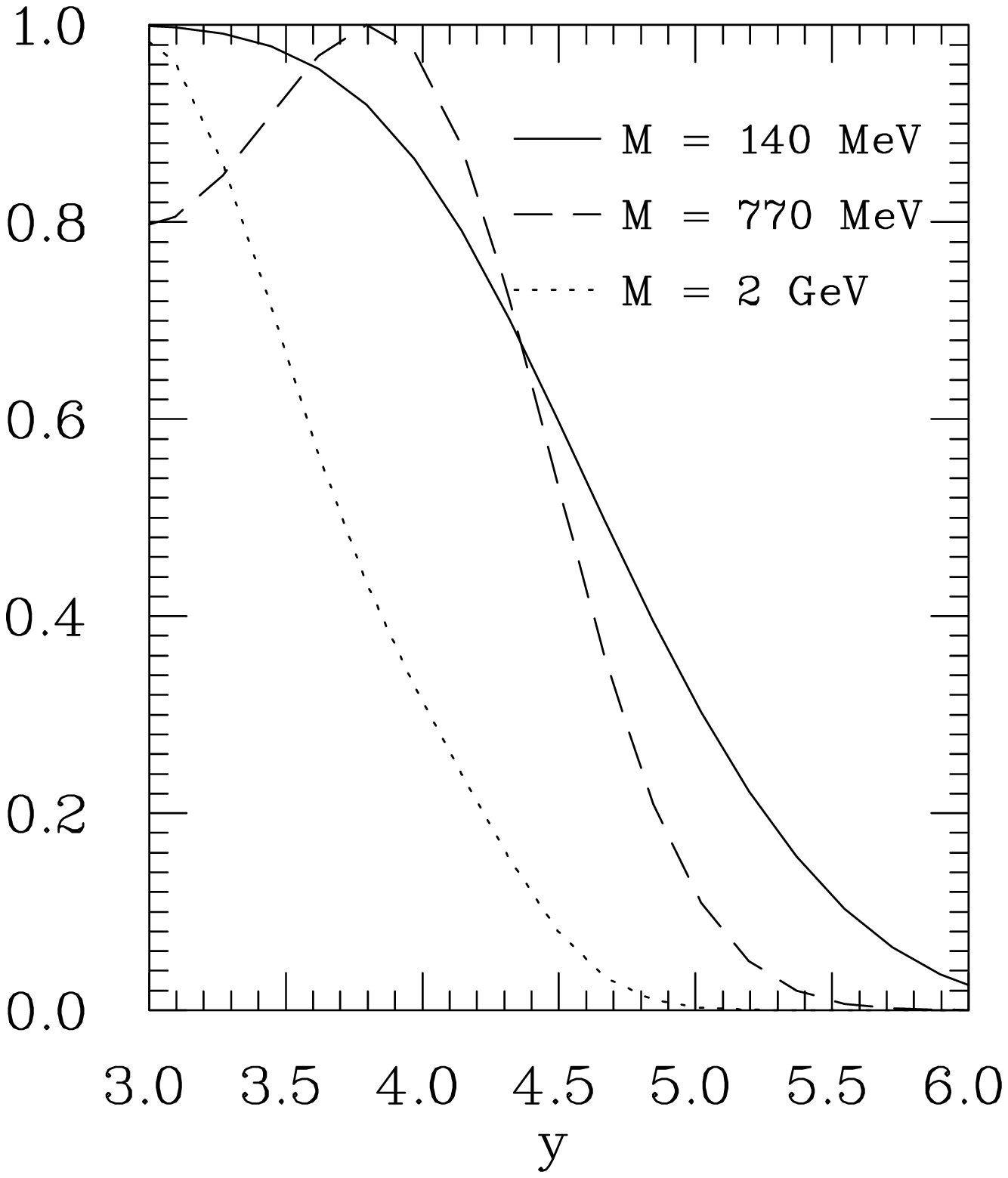}
        \hfill
   \caption{Rapidity distributions of thermal electron pairs of various
            masses for IS~1 and EoS~A scaled to the maximum value of
            the distribution.}
  \label{lepym}
  \end{minipage}
\end{center}
\end{figure}

We have found that the rapidity spectrum of dileptons may be an
observable which shows a qualitative dependence on the phase transition
temperature. In fig.~\ref{lepy10} we show the rapidity distribution of
electron pairs with invariant mass $M=770$ MeV for three different
EoSs and initial state IS~1. The shape of the distribution shows a
clear dependence on $T_c$. When EoS~H is used the distribution is
peaked at mid-rapidity, EoS~D produces a plateau and the use of EoS~A
results in a spectrum with two peaks. The difference in shape is most
prominent at $750<M<850$ MeV and vanishes at larger and lower values of
pair mass where the spectra are single-peaked for all EoSs.

Characteristic of EoS~A is the long lifetime of the fast flowing part
of the fireball compared to its center. This is a result of a small
pressure gradient in the mixed phase which does not blow the fireball
apart. Due to the long lasting mixed phase the fast flowing part stays
hot and produces a large contribution at 770 MeV invariant mass which
gives the spectrum its double peaked shape. On the other hand at
smaller masses the thermal distribution is wider. Therefore even
fast-flowing parts of the fireball contribute significantly to
mid-rapidity leptons and the resulting spectrum is single-peaked but
wide. At larger masses the emission is dominated by the hot central
region at the early stages of the evolution. The resulting spectrum is
single-peaked and narrow (see fig.~\ref{lepym}).

For EoS~D, the lifetimes of the central and the fast flowing parts
of the fireball differ less because the mixed phase is reached earlier
and its lifetime is shorter than for EoS~A. Compared to the
contribution from the central fireball the contribution from the
fast-flowing part is therefore reduced. At $M=770$ MeV the spectrum
has a plateau at mid-rapidities but at lower and higher masses the
plateau vanishes. At masses above 1.5 GeV the shape is very close to
that obtained using EoS~A indicating the dominance of the early
emission and the closeness of initial states in both cases. The
smaller emission from the fast flowing part can be seen at low masses
too where the distribution is narrower than for EoS~A.

\begin{figure}
\begin{center}
   \begin{minipage}[t]{\size}
         \epsfysize \size \epsfbox{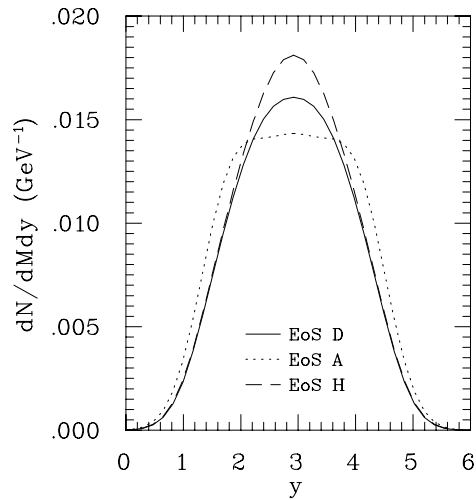}
         \hfill
   \end{minipage}
  \caption{Rapidity distributions of thermal electron pairs of mass
           770 MeV. Initial state is IS~2. No kinematic
           cuts or detector resolution has been applied.}
   \label{lepy5}
\end{center}
\end{figure}

The shape of the rapidity distribution is clearly affected by the
initial state. In fig.~\ref{lepy5} we show rapidity distributions
of $M=770$ MeV electron pairs for the initial state IS~2.
The distinctive double peaked structure for EoS~A and IS~1 has changed
into a wide plateau and EoS~D produces an almost similar shape as
EoS~H. When EoS~H is used both initial conditions result in rather
similar flows and therefore the rapidity distribution stays almost
unchanged. In the cases of EoSs A and D initial state IS~2 with
weaker flow in the central region than for IS~1 leads to longer
lifetime of the central fireball. On the other hand, compared to IS~1,
the lifetime of the mixed phase in the fast flowing parts is shortened.
As a result the double peak signal gets washed out and at masses below
770 MeV the distribution gets narrower. At masses above 1.5 GeV the
most notable difference in lepton emission for IS~1 and IS~2 is the
magnitude of the yield (see fig.~\ref{lepm}). Also the shape of the
distribution is wider for IS~2 because the lower initial temperature
cuts lepton production in the early stages of the evolution, which
contribute mainly to mid-rapidity leptons, but the small emission
from the fast-flowing parts is almost similar for both initial
conditions.

As mentioned the double peaked signal is strongest in the mass
region $750<M<850$ MeV. In our figures we have shown the spectra at
770 MeV mass because the yield is largest at the free $\rho$ mass
since no medium modifications are included in our production rates.
We want to emphasize that the doubly peaked shape is not due to some
structure in the production rates but it signals the existence of hot,
sufficiently long living and fast moving parts of the fireball. The
medium modifications of the rates can enhance or suppress this signal
at different masses by changing the relative strength of the rate in
the plasma and hadron gas in the mixed phase. Thus the region where
the distribution is double peaked may extend also to lower masses, but
the distribution will nevertheless be double peaked around $M=800$ MeV.

\section{Conclusions}

We have studied the dependence of hadron spectra on the freeze-out
temperature and the equation of state and shown that different initial
conditions of the hydrodynamical model can be used to reproduce the
data. We have also considered the dependence of electromagnetic
emission on the EoS and initial conditions when the evolution of the
fireball is constrained to be consistent with the observed hadron
spectra. 

We began our paper by studying the combined effect of freeze-out
temperature and EoS on hadronic spectra. A quantity of interest is
the mean velocity of transverse flow, $\langle v_T\rangle$, which
is not directly observable because the $p_T$ spectra depend on
temperature and --- as we want to emphasize --- the flow pattern
as well. Recently it was claimed that single particle $p_T$ spectra
of different mass are able to narrow the thermal freeze-out conditions
to $T\approx 120$ MeV~\cite{Kampfer96} because the dependence of 
$p_T$ spectra on $T_f$ and $\langle v_T\rangle$ is different for
particles of different mass. We found that this is not an unambiguous
choice but the spectra can as well be reproduced using a higher
$T\approx 140$ MeV freeze-out temperature if the EoS is stiff enough.
The difference arises for two reasons: First, we freeze-out at
constant energy density causing the baryon rich areas to decouple at
lower temperatures than the baryon poor areas, while in~\cite{Kampfer96}
a strictly constant freeze-out temperature is assumed. 
Second, in our hydrodynamical approach we get a transverse velocity
profile which is not linear as the one used in~\cite{Kampfer96}. 
For these reasons the average freeze-out temperature and the average
transverse flow velocity can not be deduced unambiguously using 
hadronic $p_T$ spectra alone but more information is needed.
HBT analysis~\cite{Wiedemann98,Schlei97} is one way to reduce this 
ambiguity.

The hadron spectra leave some freedom in choosing the initial state
of the hydrodynamic evolution. We were able to reproduce the hadron
spectra using two clearly different initial conditions for each EoS,
indicating that there is a variety of initial conditions for each EoS
which lead to acceptable hadronic spectra and still correspond to the
same thermalization timescale. In principle the difference in initial
state affects electromagnetic observables but we found that at low
values of transverse momenta the photon spectra and at low values of
invariant mass the lepton pair spectra are almost identical for both
initial conditions. On the other hand the differences increase with
the values of $k_T$ and invariant mass. For photons with 
$k_T$ \raisebox{-0.75ex}{$\stackrel{\textstyle>}{\sim}$} 5 GeV/$c$ 
and for lepton pairs with 
$M$ \raisebox{-0.75ex}{$\stackrel{\textstyle>}{\sim}$} 3 GeV 
the different initial conditions cause already almost an order of
magnitude change in the yield. The lepton pair spectrum in this
kinematic range is dominated by the Drell-Yan yield but there
is a window where the thermal yield is larger than or equal to
the Drell-Yan yield and the differences in thermal yield due to
initial state are still of the order of two (see fig.~\ref{lepm}).
These differences may be large enough to provide a possibility to
distinguish between different initial states. Since in the
intermediate mass region the thermal dilepton emission is of the
same order of magnitude as the yield of Drell-Yan pairs it can be
an important contribution to the excess observed by the NA50
collaboration~\cite{Scomparin96}.

For photons the situation is more uncertain. In the region
where differences in thermal yields are significant the total
yield may be dominated by pre-equilibrium photons masking all
the differences in the thermal yields completely. However, calculating
the yield and distribution of pre-equilibrium photons is out of scope
of this paper.

We also compared the calculated mass spectra of lepton pairs to data
measured by the CERES collaboration. The differences in the spectra
obtained using different EoSs or initial states were below the present
experimental resolution. Our production rates do not include any
in-medium effects and almost an order of magnitude enhancement is 
required to reproduce the observed excess around $M=500$ MeV
invariant mass. 

The effect of the EoS is most clearly seen in the shape of the
rapidity spectrum of $M\approx 800$ MeV lepton pairs which is very
sensitive to the lifetime of the mixed phase. The lifetime of the
mixed phase depends also on the initial state and the use of EoS~D and
IS~1 results in a very similar shape of the spectrum as the use of
EoS~A and IS~2. On the other hand these two cases can be distinguished
by their dilepton mass spectrum in the intermediate mass region where
their difference is close to an order of magnitude. We saw an
interesting double peak structure of the dilepton rapidity spectra for
a certain combination of equation of state, initial condition and
invariant mass. It would be interesting to see how it compares to an
experimental rapidity spectra.

\begin{ack}
We gratefully acknowledge helpful discussions with C.~Gale and
M.~Prakash. We thank C.~Voigt for the CERES acceptance and resolution
and V.J.~Kolhinen for calculating the Drell-Yan emission.
This work was supported by the Academy of Finland grant 27574
and the Deutsche Forschungsgemeinschaft (DFG). 

\end{ack}


\newcommand{\IJMPA}[3]{{ Int.~J.~Mod.~Phys.} {\bf A#1}, #3 (#2)}
\newcommand{\JPG}[3]{{ J.~Phys. G} {\bf {#1}}, #3 (#2)}
\newcommand{\AP}[3]{{ Ann.~Phys. (NY)} {\bf {#1}}, #3 (#2)}
\newcommand{\NPA}[3]{{ Nucl.~Phys.} {\bf A{#1}}, #3 (#2)}
\newcommand{\NPB}[3]{{ Nucl.~Phys.} {\bf B{#1}}, #3 (#2)}
\newcommand{\PLB}[3]{{ Phys.~Lett.} {\bf {#1}B}, #3 (#2)}
\newcommand{\PRv}[3]{{ Phys.~Rev.} {\bf {#1}}, #3 (#2)}
\newcommand{\PRC}[3]{{ Phys.~Rev. C} {\bf {#1}}, #3 (#2)}
\newcommand{\PRD}[3]{{ Phys.~Rev. D} {\bf {#1}}, #3 (#2)}
\newcommand{\PRL}[3]{{ Phys.~Rev.~Lett.} {\bf {#1}}, #3 (#2)}
\newcommand{\PR}[3]{{ Phys.~Rep.} {\bf {#1}}, #3 (#2)}
\newcommand{\ZPC}[3]{{ Z.~Phys. C} {\bf {#1}}, #3 (#2)}
\newcommand{\ZPA}[3]{{ Z.~Phys. A} {\bf {#1}}, #3 (#2)}
\newcommand{\JCP}[3]{{ J.~Comp.~Phys.} {\bf {#1}}, #3 (#2)}
\newcommand{\HIP}[3]{{ Heavy Ion Physics} {\bf {#1}}, #3 (#2)}
\newcommand{\EPC}[3]{{ Eur.~Phys.~J.~C} {\bf {#1}}, #3 (#2)}
\newcommand{\PrTP}[3]{{ Prog.~Theor.~Phys.} {\bf {#1}}, #3 (#2)}
\newcommand{\PrTPS}[3]{{ Prog.~Theor.~Phys.~Suppl.} {\bf {#1}}, #3 (#2)}

{}

\end{document}